\documentclass[fleqn,usenatbib]{mnras}

\usepackage{newtxtext,newtxmath}
\usepackage{xcolor}

\usepackage[T1]{fontenc}

\DeclareRobustCommand{\VAN}[3]{#2}
\let\VANthebibliography\thebibliography
\def\thebibliography{\DeclareRobustCommand{\VAN}[3]{##3}\VANthebibliography}

\usepackage{graphicx}	
\usepackage{amsmath}	
\usepackage{bbold}
\usepackage{comment}
\usepackage{bm}
\usepackage{mathtools}
\usepackage{xcolor}
\usepackage{multirow}
\usepackage{natbib}

\renewcommand{\vec}[1]{{\bm{#1}}}
\newcommand{\oper}[1]{{\bm{\mathsf{#1}}}}

\title[A robust radius for lensing perturbers]{On what scales does strong lensing robustly constrain the mass profile of low-mass perturbers?}

\author[M. Tajalli and S. Vegetti]
{M. Tajalli \thanks{E-mail: mtajalli@mpa-garching.mpg.de} and S. Vegetti
\\
Max Planck Institute for Astrophysics, Karl-Schwarzschild-Stra\ss{}e 1, 85748 Garching bei M\"unchen, Germany\\
}

\date{Accepted XXX. Received YYY; in original form ZZZ}

\pubyear{2026}

\begin{document}
\label{firstpage}
\pagerange{\pageref{firstpage}--\pageref{lastpage}}
\maketitle
\begin{abstract}

Galaxy-galaxy strong gravitational lensing is an established method to detect and characterise low-mass haloes, which are otherwise too faint to be directly observed. Recent analyses have shown empirically that there exists a characteristic radius within which the projected enclosed mass of these haloes can be robustly measured. Here we provide a physically and mathematically motivated definition of this robust radius based on the Fisher information matrix. 
We show that for any spherically symmetric smooth profile with at least two free parameters, there exists a pivot radius at which estimates of the projected mass become decorrelated from the local logarithmic slope and have minimum relative uncertainty. As a result, the projected mass within this radius can be recovered consistently by any profile that fits the data well enough.
This robust radius coincides, to leading order, with the weighted geometric mean of the arc pixel distances from the halo, where the weights depend on the perturber deflection angle, the macro-model magnification, and the source surface brightness gradient. 
Beyond leading order, corrections are of order the weighted log-width of the arc. 
The data constraining power on the local properties of the halo profile beyond amplitude is shown to depend on the radial leverage of the arc around the perturber, and on the imprint of the perturber deflection angle on high signal-to-noise-ratio pixels.
Our definition generalises to both subhaloes and line-of-sight haloes and naturally captures the important physical aspects of the problem and the effect of the data resolution and signal-to-noise ratio.

\end{abstract}

\begin{keywords}
gravitational lensing: strong -- galaxies: haloes -- cosmology: dark matter -- galaxies: structure -- methods: statistical  
\end{keywords}


\section{Introduction}

Detecting and measuring the properties of low-mass dark matter haloes via their gravitational effect on the multiple images of strongly lensed galaxies provides a key test of different dark matter models \citep[see][and references therein]{Vegetti_2024}.

To this day, three objects have been \emph{gravitationally imaged} \citep{Vegetti_2010, Vegetti_2012, Powell_2025}, two of which have since been confirmed by independent analyses with independent lens modelling codes \citep{Minor_2021, Ballard_2024, Enzi_2025, He_2025, Cao_2025}. 
Following \citet{Vegetti_2026}, we refer to these detections as object ${\cal A}$,  ${\cal H}$ and ${\cal V}$, respectively. From purely analytical modelling of the perturber profile, three further detections have been reported \citep{Hezaveh_2016,lange_2025, amvrosiadis_2026}. The first one was recently shown to be a false positive related to mismodelling of the main deflector angular structure \citep{Stacey_2025}; the remaining two have yet to be confirmed by independent pixelated analyses.

While the number of detections so far is consistent with predictions from the cold dark matter model \citep[CDM,][]{Vegetti_2014, Ritondale_2019b,Powell_2025}, given the data sensitivity, the macro-properties (e.g. virial mass and concentration) of objects ${\cal A}$,  ${\cal H}$ and ${\cal V}$ have been shown to be in tension with the theory (\citeauthor{Vegetti_2010}~\citeyear{Vegetti_2010}, \citeauthor{Minor_2021}~\citeyear{Minor_2021}, \citeauthor{Ballard_2024}~\citeyear{Ballard_2024}, \citeauthor{Minor_2025}~\citeyear{Minor_2025}, \citeauthor{Despali_2025}~\citeyear{Despali_2025},  \citeauthor{Enzi_2025}~\citeyear{Enzi_2025}, \citeauthor{Tajalli_2025}~\citeyear{Tajalli_2025}, \citeauthor{Vegetti_2026}~\citeyear{Vegetti_2026}, but see, \citeauthor{He_2025}~\citeyear{He_2025}).
However, \citet{Tajalli_2025} have shown these quantities not to be robustly measured and have proposed a different metric based on the projected mass density profile at an empirically defined robust radius. For objects ${\cal A}$ and ${\cal H}$, they found the projected mass and slope at this radius to be unusual but not in significant disagreement with the theory when compared to simulated haloes and subhaloes in the TNG50 hydrodynamical simulation. It is only when including additional constraints from the measured luminosity upper limit that this discrepancy re-emerges for object ${\cal H}$. Object ${\cal V}$ remains in strong tension even in terms of its robust quantities.

Their results clearly highlight the importance of identifying which quantities are robustly measured for a meaningful comparison with theoretical predictions. The radius proposed by \citet{Tajalli_2025} is defined as the radius around which the relative error on the inferred deflection angle reaches its minimum. At this radius, the measured cylindrical mass and slope were shown to be consistent across mass density profiles that provide a good fit to the data. A similar definition was used by \cite{Despali_2025} and \cite{Vegetti_2026}. \cite{Minor_2021} instead have proposed a different formulation based on the distance from the object centre to the point on the critical curve which is most perturbed.

In this paper, we introduce a new definition of the robust mass radius. Unlike the definition proposed by \cite{Minor_2021}, it automatically takes into account the data resolution and signal-to-noise ratio (SNR) as well as the image surface brightness gradient and magnification. Hence, by definition, it encodes how strongly the likelihood function responds to the properties of the perturber. Moreover, our new definition is formally connected to that introduced by \citet{Tajalli_2025}.

We also show that, under specific conditions, the data are not only sensitive to the mass within the robust radius but also to the deflection profile logarithmic slope and curvature, over a range of sensitive radii. This result provides a mathematical interpretation for those presented by \cite{Vegetti_2026}: although all models that adequately fit the data yield the same cylindrical mass within a given radius, they can still be distinguished and ranked by their Bayesian evidence, as data with an extended arc are sensitive to second-order to the curvature of the deflection angle profile. 

The paper is organised as follows. In Section \ref{sec:pivot_radius}, we provide a derivation of the pivot radius for generic smoothly varying mass density profiles that can be Taylor expanded about a power-law. The derivation is based on the Fisher information matrix. In Section \ref{sec:robust_radius}, we discuss the connection between the pivot radius and the robust radius. The former is the radius at which the local amplitude and logarithmic slope of the profile decorrelate, and around which the relative error on the deflection angle is minimised. The latter is defined as the radius at which different mass density profiles that provide adequate fits to the data yield the same enclosed projected mass. In Section \ref{sec:profile}, we demonstrate that under specific conditions, the strong lensing data are sensitive to second order to the curvature of the perturber deflection angle profile. We summarise our findings in Section \ref{sec:discussion}.

\section{The Pivot radius}
\label{sec:pivot_radius}

In this section, we use the Fisher information matrix to show that, for any sufficiently smooth deflection angle profile, it is possible to define a pivot radius at which the profile local amplitude and logarithmic slope are uncorrelated. We further show that, at first order, this pivot radius is common to all such profiles. 

\subsection{Fisher matrix formalism}
\label{subsec:fisher}

In the following, we refer to the surface brightness distribution of the lensed images on the lowest-redshift plane as $I_{\rm {img}}(\vec{\theta})$ and to the corresponding unlensed emission on the source plane as $I_{\rm{src}}(\vec{\beta})$. These are related to each other via the telescope blurring $\oper{B}$, the lensing operator $\oper{L}$ and noise $\vec{n}$ as follows 
\begin{equation}
    I_{\rm img}(\boldsymbol{\theta}) = \oper{B}\oper{L} I_{\rm{src}}(\boldsymbol{\beta}) + \vec{n}\,.
\end{equation}
Assuming Gaussian noise, the likelihood for the observed image $\vec{I}_{\rm{img}}$ up to an additive constant is given by
\begin{equation}
  -2\ln\mathcal{L} = \left(\mathbf{B}\mathbf{L} \bm{I}_{\rm{src}} -  \bm{I}_{\rm img}\right)^{\!T}
  \mathbf{C}^{-1}
  \left(\mathbf{B}\mathbf{L}\bm{I}_{\rm{src}} -  \bm{I}_{\rm img}\right)\,,
  \label{eq:likelihood}
\end{equation}
where $\oper{C}$ is the noise covariance matrix of the image. If the main deflector without a perturber reproduces the data well, i.e., $\mathbf{B}\mathbf{L}I_{\rm{src,\,MAP}}(\boldsymbol{\beta}) \approx  I_{\rm img}(\boldsymbol{\theta})$ for the maximum-a-posteriori (MAP) source $I_{\rm{src,\,MAP}}(\boldsymbol{\beta})$, the image residual introduced by a perturber within the lensing galaxy is related to a change in the lensing operator $\delta \vec{I}_{\rm img} = \oper{B}\,\delta\oper{L}\,\vec{I}_{\rm{src,\,MAP}}$.

Following \citet{Blandford_2001} and \citet{Koopmans_2005}, we linearise the source surface brightness around the unperturbed source position and obtain
\begin{multline}
    \delta I_{\rm img}(\boldsymbol{\theta}) \approx \oper{B}\left(\nabla_{\beta} I_{\rm src}(\boldsymbol{\beta}) \cdot \delta{\boldsymbol{\beta}}\right) = 
    \oper{B}\left(\mathbf{A}^{-1} \nabla_{\theta} \tilde{I}_{\rm img}(\boldsymbol{\theta}) \cdot \delta{\boldsymbol{\beta}}\right) =
    \\ - \oper{B}\left(\mathbf{A}^{-1} \nabla_{\theta} \tilde{I}_{\rm img}(\boldsymbol{\theta}) \cdot  \vec{\alpha}_p(\bm{\theta})\right)\,,
    \label{eq:delta_I_src} 
\end{multline}
where $\delta{\boldsymbol{\beta}}$ is the source displacement induced by the perturber deflection angle $\vec{\alpha}_{p}(\bm{\theta})$, $\oper{A}$ is the lensing Jacobian of the main deflector and $\tilde{I}_{\rm img}$ is the unconvolved image brightness $\tilde{I}_{\rm img} (\boldsymbol{\theta}) = \oper{L} I_{\rm{src}}(\boldsymbol{\beta})$.  
Under these approximations, the expected mean of the surface brightness of a lensed image that includes a perturber with parameters $\bm{\eta}$ is $\langle \bm{I}_{\rm img} \rangle = \mathbf{B}\mathbf{L}\bm{I}_{\rm{src,\,MAP}}+\langle \delta \bm{I}_{\rm img} \rangle$.

The Fisher information matrix, which quantifies how much information the data contain about a given parameter, follows from the expectation value of the negative Hessian of the log-likelihood evaluated at the fiducial perturber parameters:
\begin{equation}
    \mathcal{I} = -\left\langle
    \frac{\partial^2\ln\mathcal{L}}{\partial\eta_i\partial\eta_j}
  \right\rangle\,.
\end{equation}
Assuming uncorrelated noise in the data, the Fisher matrix elements are given as
\begin{align}
[\mathcal{I}]_{ij}
&=  \frac{\partial\langle {\delta \bm{{I}}_{\rm img}}^T \rangle}{\partial\eta_i}  \oper{C}^{-1}  \frac{\partial\langle \delta \bm{{I}}_{\rm img} \rangle}{\partial\eta_j}   \nonumber \\
&= \sum_k \frac{1}{\sigma_k^2}
    \left[\oper{B}\left(
    \nabla_\beta I_{\rm src}(\boldsymbol{\beta})
    \cdot \frac{\partial\vec{\alpha}_p}{\partial\eta_i}
    \right)\right]_k
    \left[\oper{B}\left(
    \nabla_\beta I_{\rm src}(\boldsymbol{\beta})
    \cdot \frac{\partial\vec{\alpha}_p}{\partial\eta_j}
    \right)\right]_k \nonumber \\
&= \sum_k \frac{1}{\sigma_k^2}
    \left[\oper{B}\left(
    \mathbf{A}^{-1}\nabla_\theta \tilde{I}_{\rm img}(\boldsymbol{\theta})
    \cdot \frac{\partial\vec{\alpha}_p}{\partial\eta_i}
    \right)\right]_k
    \left[\oper{B}\left(
    \mathbf{A}^{-1}\nabla_\theta \tilde{I}_{\rm img}(\boldsymbol{\theta})
    \cdot \frac{\partial\vec{\alpha}_p}{\partial\eta_j}
    \right)\right]_k \nonumber \\
&= \sum_k \frac{1}{\sigma_k^2}
    \left[\oper{B}\left(
    \mathbf{Q}\mathbf{\Lambda}^{-1}\mathbf{Q}^{T}
    \nabla_\theta \tilde{I}_{\rm img}(\boldsymbol{\theta})
    \cdot \frac{\partial\vec{\alpha}_p}{\partial\eta_i}
    \right)\right]_k \nonumber \\
&\hspace{2.2em}\times
    \left[\oper{B}\left(
    \mathbf{Q}\mathbf{\Lambda}^{-1}\mathbf{Q}^{T}
    \nabla_\theta \tilde{I}_{\rm img}(\boldsymbol{\theta})
    \cdot \frac{\partial\vec{\alpha}_p}{\partial\eta_j}
    \right)\right]_k \nonumber \\
&= \sum_k \frac{1}{\sigma_k^2}
    \left[\oper{B}\left(
    \frac{v_r w_r^{(i)}}{\lambda_r}
    + \frac{v_t w_t^{(i)}}{\lambda_t}
    \right)\right]_k
    \left[\oper{B}\left(
    \frac{v_r w_r^{(j)}}{\lambda_r}
    + \frac{v_t w_t^{(j)}}{\lambda_t}
    \right)\right]_k\,,
\label{eq:fisher}
\end{align}
where $\sigma_k^2 = [C]_{kk}$ is the noise variance at pixel $k$. Above, we have diagonalised the Jacobian matrix as $\mathbf{A} = \mathbf{Q}\mathbf{\Lambda}\mathbf{Q}^{T}$, where $\mathbf{Q}$ is an orthogonal matrix whose columns are the eigenvectors $\hat{e}_t$ and $\hat{e}_r$, and
$\mathbf{\Lambda} = \mathrm{diag}(\lambda_t, \lambda_r)$, where $\lambda_t$ and $\lambda_r$ are the tangential and radial eigenvalues, respectively. The tangential and radial projections of the deflection derivative and image gradient are defined as follows 
\begin{equation}
  w_t^{(i)}(\boldsymbol{\theta}) \equiv \hat{e}_t(\boldsymbol{\theta})\cdot\frac{\partial\vec{\alpha}_p}{\partial\eta_i}(\boldsymbol{\theta})\,, \qquad
  w_r^{(i)}(\boldsymbol{\theta}) \equiv \hat{e}_r(\boldsymbol{\theta})\cdot\frac{\partial\vec{\alpha}_p}{\partial\eta_i}(\boldsymbol{\theta})
\end{equation}
and 
\begin{equation}
  v_t(\boldsymbol{\theta}) \equiv \hat{e}_t(\boldsymbol{\theta})\cdot\nabla_\theta \tilde{I}_{\rm img}(\boldsymbol{\theta})\,, \qquad
  v_r(\boldsymbol{\theta}) \equiv \hat{e}_r(\boldsymbol{\theta})\cdot\nabla_\theta \tilde{I}_{\rm img}(\boldsymbol{\theta})\,.
\end{equation}
For a spherical perturber, we obtain the following expression
\begin{multline}
    [\mathcal{I}]_{ij} = \sum_k \frac{1}{\sigma_k^2}
    \left[\oper{B}\!\left(g_i \!\left(\frac{v_t\cos\phi}{\lambda_t}
    + \frac{v_r\sin\phi}{\lambda_r}\right)\right)\right]_k \\
    \times \left[\oper{B}\!\left(g_j \!\left(\frac{v_t\cos\phi}{\lambda_t}
    + \frac{v_r\sin\phi}{\lambda_r}\right)\right)\right]_k \,.
\label{eq:fisher_spherical}
\end{multline}
with 
\begin{equation}
    g_{i}(\theta) = \frac{\partial\alpha_p (\theta; \boldsymbol{\eta})}{\partial\eta_{i}}\,,
    \label{eq:g_i}
\end{equation}
where $\theta \equiv |\boldsymbol{\theta} - \boldsymbol{\theta}_p|$ is the distance from the centre of the perturber, $\boldsymbol{\theta}_p$, and $\phi$ is the angle between $\hat{\theta}$ and $\hat{e}_t(\boldsymbol{\theta})$. 

\begin{figure}
\centering
\includegraphics[width=1.0\linewidth]{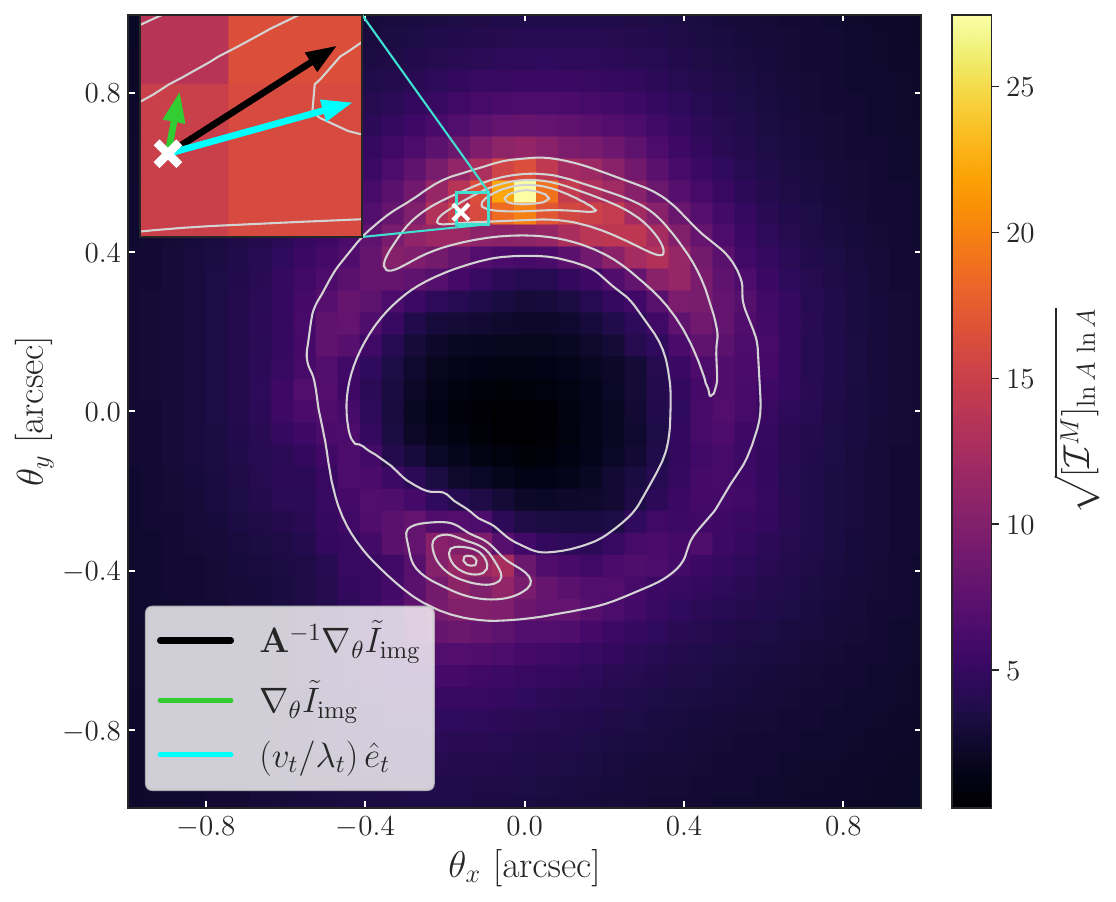}
\caption{Marginalised Fisher information for the perturber amplitude as a function of the position of an NFW subhalo with fixed mass ($M_{\rm vir}=5\times10^8 M_\odot$) and concentration ($c_{\rm vir}=120$). The lens system properties are based on Keck Adaptive-Optics (AO) K-band observations of B1938+666 (see Section \ref{sec:robust_radius}). The main-lens parameters are marginalised out in the Fisher calculation, while the source is held fixed at its MAP values. 
Contours show the surface brightness of the lensed image for the smooth (unperturbed) model. 
The cross marks an example of perturber position, with the inset showing a zoomed-in region around it, where the black, green, and cyan vectors are the effective sensitivity direction, $\mathbf{A}^{-1}\nabla_\theta I_{\rm img}$, image brightness gradient, and its tangential projection stretched by the tangential eigenvalue of the inverse Jacobian matrix, respectively.}
\label{fig:sensitivity}
\end{figure}

From the general expression of the Fisher matrix, we can conclude that constraints on a given perturber parameter are maximised when a change in that parameter deflects light along $\mathbf{A}^{-1}\nabla_\theta I_{\rm img}$, i.e., along the image gradient differentially stretched by the magnification tensor along its radial and tangential eigen-directions. In other words, the source brightness gradient and the macro-model magnification define a preferential direction: perturbations that shift light across source isophotes imprint a stronger signal on the image than those acting along them, with the effective sensitivity direction set by a combination of the magnification and source gradient; see Fig.~\ref{fig:sensitivity} for a schematic representation.

\subsection{Amplitude-slope decorrelation}
\label{subsec:decoupling}

For any smooth profile, we can locally expand the logarithmic deflection angle about an arbitrary radius $\theta_*$ as
\begin{equation}
  \ln\alpha_p(\theta;\, \boldsymbol{\eta})
  = \sum_{n\ge 0}\frac{c_n (\boldsymbol{\eta})}{n!}\,\ell^{\,n}\,, \qquad 
  c_n \equiv \left.\frac{d^n\ln\alpha_p}{d\ell^n}\right|_{\theta_*}\,,
  \label{eq:alpha_taylor}
\end{equation}
where $\ell \equiv \ln(\theta/\theta_*)$ and $\boldsymbol{\eta}$ is the set of parameters describing the perturber deflection angle profile. In this notation, the leading coefficients $c_0$, $c_1$, and $c_2$ denote, respectively, the local log-amplitude of the deflection angle, its 
logarithmic slope, and its log-parabolic curvature, which quantifies deviations from a pure power-law.

We define the pivot radius $\theta_0$ as the radius at which the inferred local log-amplitude and logarithmic slope of the deflection angle profile are uncorrelated, i.e.,
\begin{equation}
\mathrm{Cov}(c_0, c_1) \big|_{\theta_*=\theta_0} = 0\,.
\label{eq:robust_cond}
\end{equation}
This definition is not specific to any profile family and applies to any sufficiently smooth profile, regardless of its parametrisation, provided that it has at least two parameters that allow the local amplitude and slope to vary independently.
Note that the condition in Eq.~\eqref{eq:robust_cond} is equivalent to requiring that the projected enclosed mass, $M_{\rm 2D}$, and its logarithmic slope decorrelate at $\theta_0$, i.e.,
\begin{equation}
\mathrm{Cov}(\ln M_{\rm 2D}, \tfrac{d \ln M_{\rm 2D}}{d \ln \theta}) \big|_{\theta_0} = 0\,.
\end{equation}
In Section \ref{sec:robust_radius} we show how the pivot radius becomes a robust radius for models that provide an acceptable fit to the data, such that their inferred enclosed masses agree within their statistical uncertainties.

It is straightforward to see that the pivot radius has an equivalent interpretation as the radius at which the perturber deflection angle is most tightly constrained by the data.
The relative error on the perturber deflection angle at radius $\theta$ is given by
\begin{equation}
    \sigma_{\ln\alpha_p}^2(\theta) \approx \bigl[\nabla_\eta\ln\alpha_p(\theta;\boldsymbol{\eta})\bigr]^T\,\mathcal{I}^{-1}\,\nabla_\eta\ln\alpha_p(\theta;\boldsymbol{\eta})\,,
\label{eq:var_alpha}
\end{equation}
a parametrisation-independent quantity.
Here, $\mathcal{I}^{-1} = C_\eta$ is the covariance matrix of the perturber parameters.
The deflection angle variance is minimised at the radius satisfying
\begin{equation}
\begin{split}
    \frac{1}{2}\frac{d\sigma^2_{\ln\alpha_p}}{d\ln\theta} 
    &= \sum_{\eta_i,\eta_j} \frac{\partial}{\partial {\eta_i}} \left(\frac{d\ln\alpha_p}{d\ln\theta}\right)[\mathcal{I}^{-1}]_{\eta_i\eta_j}\frac{\partial}{\partial {\eta_j}}\ln\alpha_p \\
    &\equiv \mathrm{Cov}(c_0, c_1)
    = 0\,,
    \end{split}
    \label{eq:pivot_sigma_alpha}
\end{equation}
where we made use of the series expansion in Eq.~\eqref{eq:alpha_taylor} to express the radial derivative of $\sigma^2_{\ln\alpha_p}$ in terms of the local expansion coefficients, thereby recovering the decorrelation condition of Eq.~\eqref{eq:robust_cond}. The pivot radius $\theta_0$ is therefore not only the radius at which the deflection angle, or, equivalently, the projected enclosed mass, is best constrained, but also the radius at which this constraint is decorrelated with that of the local slope, or, as we show below for two-parameter profiles, with the global parameter that controls the radial dependence of the profile.

\subsubsection{From local coefficients to global halo properties}
\label{subsec:local_to_global}
Eq.~\eqref{eq:robust_cond} defines the pivot radius in terms of the local properties of the halo deflection angle profile. However, the coefficients $c_0$ and $c_1$ are derived quantities, obtained from the global parameters of the analytic model used to fit the data, which are themselves often highly degenerate with each other (e.g., the virial mass and concentration of an NFW \citep{Navarro_1996} profile). Here, we describe how the pivot radius condition is expressed in terms of these global parameters.

We first consider spherical deflection angle profiles of the form
\begin{equation}
    \alpha_p(\theta) = A\,f(\theta; s)\,,
\end{equation}
where $A$ is an amplitude parameter and $s$ determines the radial dependence through the function $f$, corresponding, for instance, to the logarithmic slope of a power-law profile or an angular scale length for the NFW and Pseudo-Jaffe \citep[PJ,][]{Munoz_2001} profiles.
Using the delta method, we can approximate the covariance between the local derived quantities $c_0$ and $c_1$ from the covariance of the global perturber parameters $\boldsymbol{\eta} = \{A,s\}$. Since $c_i$ for $i \ge 1$ depends only on the scale parameter $s$, we obtain
\begin{equation}
\mathrm{Cov}(c_0, c_1) = \frac{\partial c_1(\theta_*;s)}{\partial s} \left(\frac{[\mathcal{I}^{-1}]_{As}}{A} + [\mathcal{I}^{-1}]_{ss}  \frac{\partial \ln f(\theta_*;s)}{\partial s}\right)\,.
\label{eq:cov_robust}
\end{equation}
Although the pivot radius is uniquely determined, independently of the model parametrisation, a particular choice is especially useful for interpreting the decorrelation condition in terms of the physical parameters that characterise the global mass density profile of the halo. 
Expressing the amplitude as the deflection angle at $\theta_0$, the deflection angle reads as follows
\begin{equation}
    \alpha_p(\theta) = \alpha_p(\theta_0)\,F(\theta;s)\,,
    \label{eq:alpha_two_reparam}
\end{equation}
where $F(\theta;s) \equiv f(\theta;s)/f(\theta_0;s)$ such that $F(\theta_0;s)=1$.
With this choice of parametrisation, the pivot condition from Eq.~\eqref{eq:cov_robust} simplifies to
\begin{equation}
    [\mathcal{I}^{-1}]_{As} = 0 \,,
    \label{eq:i_as}
\end{equation}
which, for a two-parameter model, holds if the Fisher matrix is diagonal, $[\mathcal{I}]_{As}=[\mathcal{I}]_{sA}=0$; any change in the likelihood caused by varying one of the two parameters cannot be offset by adjusting the other. 

For three-parameter models described by an amplitude parameter, a scale parameter $s_1$, and an additional shape parameter $s_2$,
we can similarly parametrise the profile as 
\begin{equation}
    \alpha_p(\theta) = A\,F(\theta; s_1,s_2)\,,
    \label{eq:alpha_three_reparam}
\end{equation}
where $F(\theta_0;s_1,s_2)=1$ and $A \equiv \alpha_p(\theta_0)$.
Imposing the pivot condition (Eq.~\ref{eq:robust_cond}) then yields
\begin{equation}
[\mathcal{I}^{-1}]_{As_1} \frac{\partial c_1}{\partial {s_1}}+ [\mathcal{I}^{-1}]_{As_2} \frac{\partial c_1}{\partial {s_2}} = 0\,.
\label{eq:pivot_three_param}
\end{equation}
Thus, while the pivot radius is well-defined for three-parameter models, the amplitude remains correlated with both the scale and shape parameters at this radius. 
In general, the existence of a radius at which the amplitude decorrelates from both the scale and shape parameters would imply that $[\mathcal{I}^{-1}]_{As_1}=0$ and $[\mathcal{I}^{-1}]_{As_2}=0$ hold simultaneously, which requires a fine-tuning between the arc geometry and the halo profile.

\subsection{Analytical approximation}
\label{subsec:analytic}

We now assume that the derivatives of the deflection profile with respect to the perturber parameters, $g_{i}$ (Eq.~\ref{eq:g_i}), vary slowly over the size of the point spread function (PSF).
With this approximation, the Fisher matrix simplifies to 
\begin{equation}
    [\mathcal{I}]_{ij} = \sum_k W_k\,g_i({\theta}_k) g_j({\theta}_k)\,,
    \label{eq:fisher_spherical_approx}
\end{equation}
with the following weights for a spherical perturber, 
\begin{equation}
  W_k \equiv \frac{1}{\sigma_k^2}\left[\oper{B}\!\left(\frac{v_t\cos\phi}{\lambda_t} + \frac{v_r\sin\phi}{\lambda_r}\right)\right]_k^{2}\,,
\end{equation}
allowing for an analytic treatment. In the following, we derive useful analytical expressions for the Fisher matrix and the pivot radius of a slowly varying halo profile described by two- and three-parameter models separately.

\subsubsection{Two-parameter models}
\label{subsec:two_param}

As we saw in Section \ref{subsec:decoupling}, for two-parameter profiles the pivot condition is equivalent to setting the off-diagonal term of the Fisher matrix to zero,
\begin{equation}
  [\mathcal{I}]_{As} = [\mathcal{I}]_{sA} = \sum_k W_k\,g_A({\theta}_k)g_s({\theta}_k) = \sum_k \widetilde{W}_k\,\frac{g_s({\theta}_k)}{g_A({\theta}_k)} = 0\,,
\end{equation}
where $\widetilde{W}_k \equiv W_k \cdot {g_A({\theta}_k)}^2 \geq 0$, and we have expressed the deflection angle model using the parametrisation of Eq.~\eqref{eq:alpha_two_reparam}.
The condition $[\mathcal{I}]_{As}=0$ is a balance condition, contributions from pixels where $\frac{g_s}{g_A}>0$ must cancel those where $\frac{g_s}{g_A}<0$. In other words, the ratio $\frac{g_s}{g_A}$ must change sign across the lensed arc. 

For the special case of a power-law profile, we can express the deflection angle as follows
\begin{equation}
  \alpha_{\rm PL}(\theta) \propto \frac{M(\theta_0)}{\theta_0}\left(\frac{\theta}{\theta_0}\right)^{1-\gamma}\,,
  \label{eq:alpha_pl}
\end{equation}
where $\gamma$ is the slope of the projected mass density profile and, for notational simplicity, $M$ denotes the projected enclosed mass $M_{\rm 2D}$. The condition of a diagonal Fisher matrix then becomes
\begin{equation}
[\mathcal{I}]_{M\gamma} = -\frac{1}{M(\theta_0)}\sum_k W_k\,\alpha_p^2(\theta_k)\ln\left(\frac{\theta_k}{\theta_0}\right) = 0\,.
\label{eq:offdiag_pl}
\end{equation}
Solving for $\theta_0$ leads to
\begin{equation}
\ln \theta_0 = \frac{\sum_k W_k\,\alpha_p^2(\theta_k)\,\ln \theta_k}{\sum_k W_k\,\alpha_p^2(\theta_k)}\,,
\label{eq:robust}
\end{equation}
and specifically in terms of the power-law parameters,
\begin{equation}
\ln \theta_0  =
\frac{\sum_k W_k\,\theta_k^{2(1-\gamma)}\ln \theta_k}{\sum_k W_k\,\theta_k^{2(1-\gamma)}}\,.
\label{eq:robust_PL}
\end{equation}
Hence, for a spherical power-law profile, the pivot radius is the weighted geometric mean of the pixel distances from the perturber. 
At this radius, the covariance matrix takes the form
\begin{equation}
  \mathcal{I}^{-1}\big|_{\theta_0} = 
  \begin{pmatrix} 
     \dfrac{M(\theta_0)^2}{\displaystyle\sum_k W_k\,\alpha_p^2(\theta_k)} & 0 \\[2ex] 
    0 &  \dfrac{1}{\displaystyle\sum_k W_k\,\alpha_p^2(\theta_k)\,\ln^2(\theta_k/\theta_0)} 
  \end{pmatrix}\,,
\end{equation}
with diagonal entries $\sigma_M^2$ and $\sigma_\gamma^2$, respectively. The ratio between the relative error on the perturber enclosed mass and the error on its slope at the pivot radius is then
\begin{equation}
    \frac{\sigma_M/M}{\sigma_\gamma} = 
    \sqrt{\frac{\sum_k w_k \ln^2(\theta_k/\theta_0) }{\sum_k w_k}}\,,
    \label{eq:sigmaM_sigmagamma}
\end{equation}
where $w_k \equiv W_k\,\alpha_p^2(\theta_k)$. The relative error on the perturber enclosed mass scales inversely with the total pixel weight, and is therefore smallest when the highest-SNR pixels of the lensed image overlap with the region where the perturber deflection angle is largest.
The slope variance additionally depends on the log-width of the pixel distribution around the pivot radius, so constraining the slope requires the high-SNR pixels of the arc to span a range of radii around $\theta_0$. 

For profile families beyond power-law (such as the NFW profile), imposing the condition $[\mathcal{I}]_{As} = 0$ gives
\begin{equation}
 \frac{\partial c_1}{\partial s}\sum_k w_k\,\ell_k
  + \frac{1}{2}\frac{\partial c_2}{\partial s}\sum_k w_k\,\ell_k^2
  + \mathcal{O}\!\left(\sum_k w_k\,\ell_k^3\right) = 0\,,
  \label{eq:fisher_expanded}
\end{equation}
where $\ell_k\equiv\ln(\theta_k/\theta_0)$. The first term is the power-law contribution, identical in structure to Eq.~\eqref{eq:offdiag_pl}. 
The second-order correction depends on the weighted squared log-width of the pixel distribution around the perturber and how the local curvature of the profile responds to changes in the global scale parameter. The curvature correction therefore results in a shift in the leading-order estimate for the pivot radius (Eq. \ref{eq:robust}) by
\begin{equation}
  \Delta \ln \theta_0 \approx \frac{\partial c_2/\partial s}{2\,\partial c_1/\partial s} \frac{\sum_k w_k\,\ell_k^2}{\sum_k w_k}\,.
  \label{eq:pivot_shift}
\end{equation}

Through propagating the uncertainties in the global model parameters, we can then estimate the covariance matrix of the local expansion coefficients of the perturber profile at the pivot radius,
\begin{equation}
  C_{c({\bm{\eta}})} = J C_\eta J^T\,,  
\end{equation}
where $J$ is the Jacobian $[J]_{ij} = \frac{\partial c_i}{\partial \eta_j}$ evaluated at $\theta_0$.
This gives
\begin{equation}
  \sigma^2_{c_0}=\frac{[\mathcal{I}^{-1}]_{AA}}{A^2}\,, \qquad  \sigma^2_{c_n}=[\mathcal{I}^{-1}]_{ss}{\left(\frac{\partial c_n (s;\theta_0)}{\partial s}\right)}^2 \quad (n\ge1)\,,
\end{equation}
with the following correlation coefficients at $\theta_0$:
\begin{equation}
  |r_{c_mc_n}|=1\,, \quad r_{c_0c_n}=0 \quad (m,n \ge 1)\,.
\end{equation}
Hence, the pivot condition that decorrelates the amplitude from the logarithmic slope also decorrelates it from every higher-order expansion coefficient (as can be inferred from Eq.~\ref{eq:i_as}), while those coefficients remain perfectly correlated with one another. The error on the higher-order local coefficients at the pivot radius depends on how sensitive each coefficient is to the global parameter $s$ of a given model.

\subsubsection{Three-parameter models}
\label{subsec:three_param}

We now turn our attention to three-parameter profiles, parametrised as in Eq.~\eqref{eq:alpha_three_reparam}.
The non-singular three-parameter models such as the Einasto \citep{Einasto_1965} and generalized NFW \citep[gNFW,][]{Zhao_1996} profiles have been shown to fit the stacked density profiles of dark matter haloes in $\Lambda$CDM simulations better than the two-parameter NFW model \citep{Navarro_2004, Merritt_2006, Gao_2008, Hayashi_2008, Navarro_2010}. 

From Eq.~\eqref{eq:pivot_three_param}, it is clear that locating the pivot radius of these profiles depends on how the scale and shape parameters control the local slope; it is therefore instructive to ask whether strong lensing data allow us to distinguish between these two parameters in the first place. To this end, we compute the shape block $\mathcal{S}$ of the $3\times3$ Fisher matrix, conditioned on the amplitude parameter $A$
\begin{equation}
    [\mathcal{S}]_{ij} = [\mathcal{I}]_{s_i s_j} \approx \sum_{m,n \ge 1} \frac{\bigl({\partial c_m}/{\partial{s_i}}\bigr) \bigl({\partial c_n}/{\partial{s_j}}\bigr)}
    {m!\,n!}\,\mu_{m+n}\,,
\end{equation}
for $i, j \in \{1, 2\}$, and we have introduced the moments of the logarithmic pixel distances from the pivot radius,
\begin{equation}
    \mu_n \equiv \sum_k w_k\,\ell_{k}^n\,.
\end{equation}
Truncating the Taylor expansion up to the $N$-th order term, we can then decompose $\mathcal{S}$ into an $N \times 2$ Jacobian matrix ${[J^\prime]_{mj}} = \frac{\partial c_m}{\partial s_j}$, and an $N \times N$ matrix $D$ that is data-dependent,
\begin{equation}
    \mathcal{S} = {J^\prime}^T D {J^\prime} \,,
    \qquad
    [D]_{mn} = \frac{\mu_{m+n}}{m!n!} \,,
\end{equation}
with indices $m$ and $n$ starting from 1. Keeping terms in the expansion up to the local slope, the shape block becomes, to first order,
\begin{equation}
\mathcal{S}^{(1)} = \mu_2 {J^{\prime(1)}}^T{J^{\prime(1)}}\,.
\label{eq:s_block}
\end{equation}
At this order, the shape block is singular, $\det\mathcal{S}^{(1)}=0$. The singularity of $\mathcal{S}^{(1)}$ implies that the log-likelihood is flat along one direction in the $(s_1, s_2)$ plane, so that $s_1$ and $s_2$ cannot be determined independently and remain degenerate with each other. Breaking the degeneracy between the scale and shape parameters therefore requires information beyond the local slope of the profile at $\theta_0$, encoded in the higher moments.

Including the curvature of the profile, the determinant of the shape block becomes 
\begin{equation}
\det \mathcal{S}^{(2)} = \frac{1}{4} \left(\mu_2\mu_4 - \mu^2_3\right) (\det J^{\prime(2)})^2 \,.
\label{eq:shape_curv}
\end{equation}
The determinant of $J^{\prime(2)}$ depends only on the profile family's parametrisation and is in general non-zero, while $\mu_2\mu_4 - \mu^2_3$ is fixed by the geometry of the lensed arc and the pixel weights about $\theta_0$. As $\mu_2\mu_4 - \mu^2_3 \to 0$, the covariance ellipse becomes increasingly stretched and data ability to separate $s_1$ from $s_2$ decreases. 

An immediate consequence of the results above is that if the data lack sufficient radial leverage, i.e., the high signal-to-noise pixels do not span a wide enough range of radii about $\theta_0$ (quantified through the data-dependent term in Eq.~\ref{eq:shape_curv}), the profile features beyond the local slope are not constrained, and fitting such data with a three-parameter model leaves the scale and shape parameters degenerate.

\subsection{Field haloes}

Up to this point, we have only considered lensing perturbers at the same redshift as the main deflector. 
Here, we show how the pivot radius can similarly be defined for line-of-sight perturbers, by imposing the amplitude-slope decorrelation condition on their own redshift planes.
The multi-plane lens equation for a system with $N$ deflectors at redshifts $z_i$, ordered such that $i > j$ implies $z_i > z_j$, is given by 
\begin{equation}
\label{eq:mp_lens}
\boldsymbol{\theta}^{(j)} = \boldsymbol{\theta}^{(1)} - \sum_{i=1}^{j-1} \beta_{ij} \boldsymbol{\alpha}_i(\boldsymbol{\theta}^{(i)}),
\qquad
\beta_{ij} = \frac{D_{ij} D_s}{D_j D_{is}}\,,
\end{equation}
where $\boldsymbol{\theta}^{(i)}$ is the angular position on the $i$-th lens plane and $\boldsymbol{\alpha}_i(\boldsymbol{\theta}^{(i)})$ is the deflection angle of a light ray that traverses the $i$-th lens plane at $\boldsymbol{\theta}^{(i)}$. $D_{ij}$ and $D_{i}$ are the angular diameter distances between the $i$-th and $j$-th redshift planes, and between the observer and the $i$-th plane, respectively, with index $s = N+1$ corresponding to the source plane. For notational simplicity, we denote the image plane position $\boldsymbol{\theta}^{(1)}$ by $\boldsymbol{\theta}$ in what follows. The mapping from the image plane onto the source plane is then
\begin{equation}
\label{eq:img_to_src}
\boldsymbol{\theta}^{(s)} = \boldsymbol{\theta} - \sum_{i=1}^{N}  \boldsymbol{\alpha}_i(\boldsymbol{\theta}^{(i)})\, \equiv \boldsymbol{\theta} - \boldsymbol{\alpha}_{\rm eff} (\boldsymbol{\theta})\,,
\end{equation}
where $\boldsymbol{\alpha}_{\rm eff}$ is the total effective deflection angle on the image plane.
The effective deflection of the line-of-sight perturber, $\boldsymbol{\delta\alpha}_{\rm eff}$, then follows by subtracting the main lens deflection $\boldsymbol{\alpha}_L$,
\begin{equation}
\boldsymbol{\delta\alpha}_{\rm eff}(\boldsymbol{\theta}) \equiv \boldsymbol{\alpha}_{\rm eff}(\boldsymbol{\theta}) - \boldsymbol{\alpha}_L(\boldsymbol{\theta})\,.
\end{equation}
For a foreground perturber, the light rays intersect the main-lens plane at a position shifted by the perturber deflection, and the effective deflection is
\begin{equation}
\boldsymbol{\delta\alpha}_{\rm eff}(\boldsymbol{\theta}) = \boldsymbol{\alpha}_p(\boldsymbol{\theta}) + \boldsymbol{\alpha}_L\bigl(\boldsymbol{\theta} - \beta_{pL}\boldsymbol{\alpha}_p(\boldsymbol{\theta})\bigr) - \boldsymbol{\alpha}_L(\boldsymbol{\theta})\,,
\end{equation}
where the subscripts $p$ and $L$ in the distance-ratio factor $\beta$ denote the perturber and main-lens redshift planes, respectively.
Since the perturber deflection is small compared to that of the main-lens galaxy, it shifts the light rays by much smaller scales than those over which the main-lens deflection varies. Taylor expanding $\boldsymbol{\alpha}_L$ around $\boldsymbol{\theta}$ and substituting the main-lens Jacobian, $\mathbf{A}$, yields
\begin{equation}
\boldsymbol{\delta\alpha}_{\rm eff}(\boldsymbol{\theta}) = \bigl[(1-\beta_{pL})\mathbf{I} + \beta_{pL}\mathbf{A}(\boldsymbol{\theta})\bigr]\,\boldsymbol{\alpha}_p(\boldsymbol{\theta})+ \mathcal{O}(\alpha_p^2(\boldsymbol{\theta}))\,.
\end{equation}
For a background perturber, the light rays reach the perturber plane after being deflected by the main-lens, at the shifted position $\boldsymbol{\theta} - \beta_{Lp}\boldsymbol{\alpha}_L(\boldsymbol{\theta})$. The effective deflection is then
\begin{equation}
    \boldsymbol{\delta\alpha}_{\rm eff} = \boldsymbol{\alpha}_p(\boldsymbol{\theta} - \beta_{Lp}\boldsymbol{\alpha}_L(\boldsymbol{\theta}))\,.
\end{equation}
Assuming spherical symmetry for the perturber deflection angle on its own plane, the effective deflection of a line-of-sight perturber at redshift $z_p$ factorises as
\begin{equation}
\label{eq:alpha_eff_los_sph}
\boldsymbol{\delta\alpha}_{\rm eff}(\boldsymbol{\theta})
=
\alpha_p(\theta^\prime;\boldsymbol{\eta})\,\boldsymbol{u}\, ,
\end{equation}
where 
\begin{equation}
\boldsymbol u =
\begin{cases}
\bigl[(1-\beta_{pL})\mathbf{I} + \beta_{pL}\mathbf{A}(\boldsymbol{\theta})\bigr]\,\hat{\boldsymbol{\theta}}^\prime,
& z_p < z_{\rm lens}, \\[1ex]
\hat{\boldsymbol{\theta}}^\prime,
& z_p > z_{\rm lens} \,,
\end{cases}
\label{eq:u_los_sph}
\end{equation}
and $\theta^\prime = |\boldsymbol{\theta}^\prime - \boldsymbol{\theta}_p^\prime|$ and $\hat{\boldsymbol{\theta}}^\prime$ are the distance and unit vector from the perturber centre on its own redshift plane, $\boldsymbol{\theta}_p^\prime$, to the position onto which $\boldsymbol{\theta}$ maps on the perturber plane via Eq.~\eqref{eq:mp_lens}. 
The image residual induced by the low-mass line-of-sight halo can then be written to first order as follows
\begin{equation}
\delta I_{\rm img}(\boldsymbol{\theta}) \approx 
- \oper{B}\left(\mathbf{A}^{-1}\nabla_{\theta} \tilde{I}_{\rm img}(\boldsymbol{\theta}) \cdot  \boldsymbol{\delta\alpha}_{\rm eff}(\boldsymbol{\theta})\right)\,,
\end{equation}
which, using the slowly varying profile approximation of Section \ref{subsec:analytic}, results in the Fisher matrix 
\begin{equation}
    [\mathcal{I}]_{ij}(z_p) =
    \sum_k W_k(z_p)\,g_i({\theta_k^\prime})\,g_j({\theta_k^\prime})\,,
\end{equation}
where the pixel weights depend on the perturber redshift and are given by
\begin{equation}
W_k(z_p) \equiv \frac{1}{\sigma_k^2}\left[\oper{B}\!\left(\mathbf{A}^{-1}\nabla_{\theta}\tilde{I}_{\rm img}\cdot\boldsymbol{u}\right)\right]_k^{2}\,.
\label{eq:Wk_los}
\end{equation}
Here, $g_i$ is evaluated on the perturber redshift plane. 
The pivot radius of the line-of-sight halo then follows by imposing the amplitude-slope decorrelation condition on the perturber plane, as for the single-plane subhaloes. For the power-law model, it is the redshift-dependent weighted geometric mean of the distances from the perturber centre on its own plane, i.e.,
\begin{equation}
    \ln\theta'_0(z_p) = \frac{
    \sum_k W_k(z_p)\,\alpha_p^2(\theta'_k)\,\ln \theta'_k}
    {\sum_k W_k(z_p)\,\alpha_p^2(\theta'_k)} \,,
    \label{eq:robust_los}
\end{equation}
The pixel weight $W_k(z_p)$ depends on the multi-plane geometry, and therefore the same lensed arc weights the image pixels differently for perturbers at different redshifts. 
For a foreground perturber $\boldsymbol{\theta'_k} = \boldsymbol{\theta_k}$, and the distances entering the geometric mean are identical to those on the image plane. In this case, therefore, the pivot radius shifts from the single-plane case only through the factor $\boldsymbol{u}$ in $W_k(z_p)$. 
For a background perturber, the main-lens deflection shifts the image pixel positions on the perturber plane.
The mapping shrinks the distances between the centres of adjacent pixels along the tangential arc, so the same arc probes the perturber at smaller scales and the pivot radius decreases as the perturber moves further away towards the source. 

\section{The robust radius for mass estimates}
\label{sec:robust_radius}

The Fisher information matrix provides a local, model-dependent measure of how strongly the likelihood responds to changes in the model parameters. For a given parametrisation of the perturber mass profile, it identifies a pivot radius at which the model parameters may be uncorrelated, depending on the number and choice of parameters. Since the Fisher matrix depends on the chosen model, the pivot radius is, in general, also model dependent, and there is no a priori reason for different fitted families to share the same pivot radius. 

As shown in Section \ref{sec:pivot_radius}, a useful simplification arises for mass profiles that can be accurately represented by a Taylor expansion about a power-law profile over the radial range probed by the observations. To first order, such profiles are completely characterised by their local amplitude and logarithmic slope. 
The Fisher matrix then naturally decouples these two quantities at a pivot radius.
As a result, all models that admit the same local power-law description share the same pivot radius to first order, despite their different global functional forms. This condition is expected to be quite general. Any sufficiently smooth density profile is locally well approximated by a power-law over a small enough radial interval, implying that most lens models commonly used in practice satisfy this assumption. In more complex systems, different radial ranges may admit distinct local power-law descriptions. 

At the same time, a few studies \citep{Despali_2025, Tajalli_2025,Vegetti_2026} have empirically identified the existence of a robust radius, that is, a radius at which all models that provide an adequate fit to the data recover essentially the same projected enclosed mass, despite differences in their parametrisations. This is an observational result and is not guaranteed by the Fisher analysis alone.

There is, in principle, no a priori reason why the robust radius should coincide with the Fisher pivot radius. Indeed, models that provide a bad fit to the data may also have pivot radii at which their parameters decorrelate, and the error on their deflection angle is minimised. For models that are acceptable fits, the connection between the pivot radius and the robust radius arises because the combination of the true perturber profile, the lensing geometry, and the observational setup, including the source structure, noise properties, and point spread function, imprints a preferred local power-law description onto the likelihood. Since strong lensing observations probe a finite range of radii around the perturber, the data primarily constrain the tangent power-law approximation to the true mass profile over this region. This local approximation is naturally anchored at the characteristic radius $\theta_{\rm ch}$ set by the data and true perturber profile.

Any sufficiently flexible model capable of reproducing this local power-law behavior will therefore converge to nearly the same Fisher pivot radius $\theta^{m}_{\rm p}$, close to $\theta_{\rm ch}$, while simultaneously inferring nearly the same enclosed mass $M(<\theta^{m}_{\rm p})\sim M(<\theta_{\rm ch})$ and provide to first order a good fit to the data. Under these conditions, the common pivot radius of the models coincides with the characteristic radius of the true profile. The pivot radius thus becomes a robust lensing radius at which the enclosed mass is nearly unaffected by the adopted parametrisation.

As an example, we now show how the pivot radii of different models fitted to the data relate to the robust radius. To this end, we make use of mock lensing data with properties based on the strong gravitational lens system JVAS B1938+666 ($z_{\rm{lens}} = 0.881$ and $z_{\rm{s}} = 2.059$) as observed with the AO system on the Keck II telescope. From these observations, the detection of a low-mass dark matter halo has been reported \citep[object $\cal{A}$,][]{Vegetti_2012}. The pixel scale of the NIRC2 camera is $0.01$ arcsec, with a PSF full-width-half-maximum (FWHM) of $0.07$ arcsec. The mock data includes a perturber with an NFW profile of virial mass $M_{\rm vir}=5\times10^8 M_\odot$ and concentration $c_{\rm vir}=120$, consistent with the properties of the detected object when modelled as a substructure of the main lens galaxy. As shown by \cite{Tajalli_2025}, the detected object in this system is in fact a foreground line-of-sight halo; here, however, we place the perturber at the redshift of the main lens for illustrative purposes. The final mock data includes Poisson noise from the source surface brightness and Gaussian noise from the sky background.
We model the mock data with the \textsc{pronto} lens modelling code, which is based on a semi-linear inversion method within a hierarchical Bayesian framework and with an adaptive background source \citep{Vegetti_2009,Rybak_2015,Rizzo_2018,Ritondale_2019a,Powell_2021,Ndiritu_2025}. The main lens galaxy is modelled as an elliptical power-law mass distribution with external shear. For the low-mass halo, we consider an NFW profile with free concentration, an NFW profile with concentration set by the median of the redshift-dependent mass–concentration relation of \citet{Duffy_2008}, and a spherical power-law (PL). The projected halo position is free in all cases. The remaining free parameters of the perturber are $M_{\rm vir}$ and $c_{\rm vir}$ for the free-concentration NFW model, $M_{\rm vir}$ for the NFW model with fixed concentration, and the density normalisation and the logarithmic slope for the PL. The Bayesian evidence for each model, i.e., the marginal likelihood, is computed using the \textsc{MultiNest} importance nested-sampling algorithm \citep{feroz13}. We find that the free-concentration NFW and PL models provide comparably good fits to the data, with a difference in log-evidence of only $\sim 1$ in favour of the former. The fixed-concentration NFW profile is instead strongly disfavoured, with a log-evidence lower by $\sim 15$ than that of the free-concentration model.

Before computing the pivot radius of our best-fit models, we recall that the definition introduced in Section \ref{sec:pivot_radius} assumes that the smooth macro-model and its corresponding MAP source surface brightness distribution adequately represent the data. Under this assumption, the Fisher information matrix is computed at fixed macro-model and source parameters, and the pivot radius follows from the conditional covariance of the perturber parameters. Consequently, it does not account for degeneracies between the parameters of the perturber, the main lens, and the source. Here instead, we compute the marginalised pivot radius, which includes these degeneracies.
The full Fisher matrix that includes the macro-model and source parameters, $\bm{\psi}$ (i.e., the parameters jointly fitted with the perturber), can be written in terms of the sub-matrices that contain these parameters
\begin{equation}
  \mathcal{I}^{\rm full} = 
  \begin{pmatrix} 
    \mathcal{I}_{\bm{\eta}\bm{\eta}} & \mathcal{I}_{\boldsymbol{\eta}\boldsymbol{\psi}} \\[0ex] 
    \mathcal{I}_{\boldsymbol{\psi}\boldsymbol{\eta}} &  
    \mathcal{I}_{\boldsymbol{\psi}\boldsymbol{\psi}}
  \end{pmatrix}\,,
\end{equation}
where the $\mathcal{I}_{\bm{\eta}\bm{\eta}}$ block is the conditional Fisher matrix built in the same way as in the previous sections.
The marginal covariance matrix of the perturber parameters is given by the inverse of the marginalised Fisher matrix, $(\mathcal{I}_{\bm{\eta}\bm{\eta}}^{M})^{-1}$, where $\mathcal{I}_{\bm{\eta}\bm{\eta}}^{M}$ can be computed via the Schur complement \citep[e.g.][]{zhang05}
\begin{equation}
  \mathcal{I}_{\bm{\eta}\bm{\eta}}^{M} = \mathcal{I}_{\bm{\eta}\bm{\eta}} - \mathcal{I}_{\bm{\eta}\bm{\psi}}\,\mathcal{I}^{-1}_{\bm{\psi}\bm{\psi}}\,\mathcal{I}_{\bm{\psi}\bm{\eta}}\,.
  \label{eq:schur}
\end{equation}
Our definition of the pivot radius extends naturally to the marginalised case, by imposing the same amplitude-slope decorrelation condition with the covariance now given by $(\mathcal{I}_{\bm{\eta}\bm{\eta}}^{M})^{-1}$. For the two-parameter families considered here and the parametrisation of Eq.~\eqref{eq:alpha_two_reparam}, the pivot condition reads $[\mathcal{I}_{\bm{\eta}\bm{\eta}}^{M}]_{As} = 0$.

\begin{figure}
\centering
\includegraphics[width=1.0\linewidth]{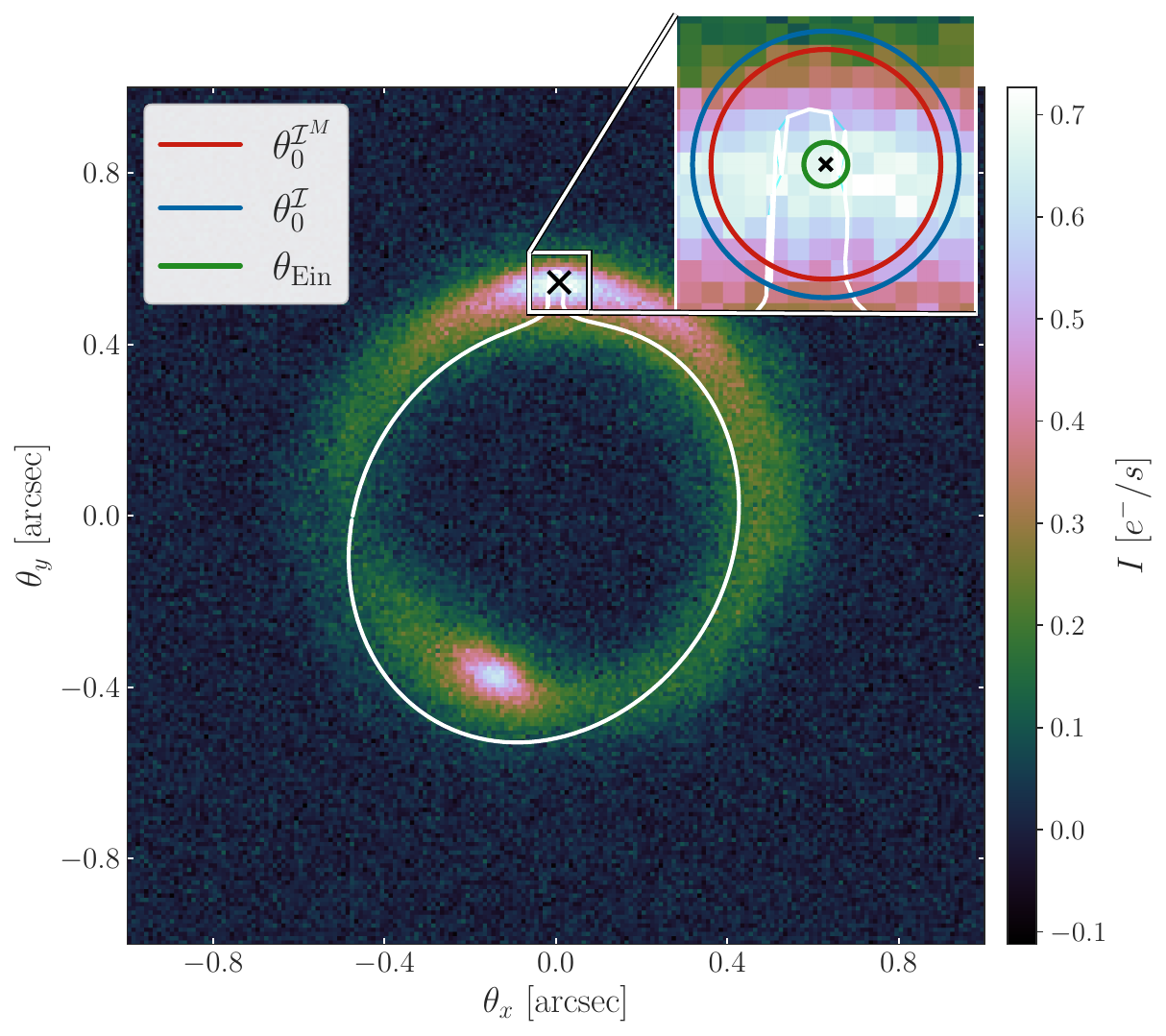}
\caption{Simulated mock data for the gravitational lens system B1938+666. The cross marks the projected position of the subhalo, and the white curve shows the perturbed critical curve. The inset shows a zoomed-in region around the perturber. The blue circle is the pivot radius of the NFW subhalo, $\theta_0^{\mathcal{I}}$, obtained from the conditional Fisher matrix, while the red circle is the pivot radius of the fitted NFW model, $\theta_0^{\mathcal{I}^M}$, computed from the nested sampling chains and therefore includes the degeneracies between the perturber, macro-model and background source. The green circle indicates the Einstein radius of the subhalo, $\theta_{\rm Ein}$, as inferred from the power-law model.}
\label{fig:pivots}
\end{figure}

The equivalence between this condition and the minimum of $\sigma_{\ln\alpha_p}^2(\theta)$ likewise holds after marginalisation. From Eq.~\eqref{eq:schur} it follows that the correction term, $\mathcal{I}_{\bm{\eta}\bm{\psi}}\,\mathcal{I}^{-1}_{\bm{\psi}\bm{\psi}}\,\mathcal{I}_{\bm{\psi}\bm{\eta}}$, can shift the pivot radius relative to the one computed from the conditional Fisher matrix, when the perturber parameters are fitted jointly with the macro-model and the source. The correction is expected to be small for perturbers that produce sharp, localised distortions at positions of high effective sensitivity on the image plane.
In Fig. \ref{fig:sensitivity} we show a colour map of the marginalised Fisher information for the perturber amplitude (with fixed profile shape) as a function of position on the image plane for our mock data, where the marginalisation is over the main lens parameters only, via the Schur complement of Eq. \eqref{eq:schur}. Note that this map is calculated through a forward-modelling approach, i.e., by evaluating the derivatives of the model with respect to the parameters at their fiducial values, without fitting the perturber.
As can be seen, the sensitivity of the data to a perturber of fixed profile shape is not uniform across the image plane and depends strongly on where the perturber is located.

In practice, we compute the pivot radii of our models directly from the nested-sampling chains, by finding the radius at which the sample variance of $\ln\alpha_p(\theta)$ is minimised. 
Note that the marginalised covariance computed from the Fisher matrix is an approximation to the empirical covariance of the posterior samples under the assumption of a Gaussian likelihood.
Fig.~\ref{fig:pivots} compares the pivot radius inferred from the posterior samples of the NFW model with free concentration with that predicted by the conditional Fisher matrix of the injected halo, evaluated before any fitting. The two radii differ by $\Delta \theta = 8$ milli-arcsec (well below the data resolution scale) and are both considerably larger than the Einstein radius inferred for the halo when it is modelled as a PL. 

\begin{figure}
\centering
\includegraphics[width=1.0\linewidth]{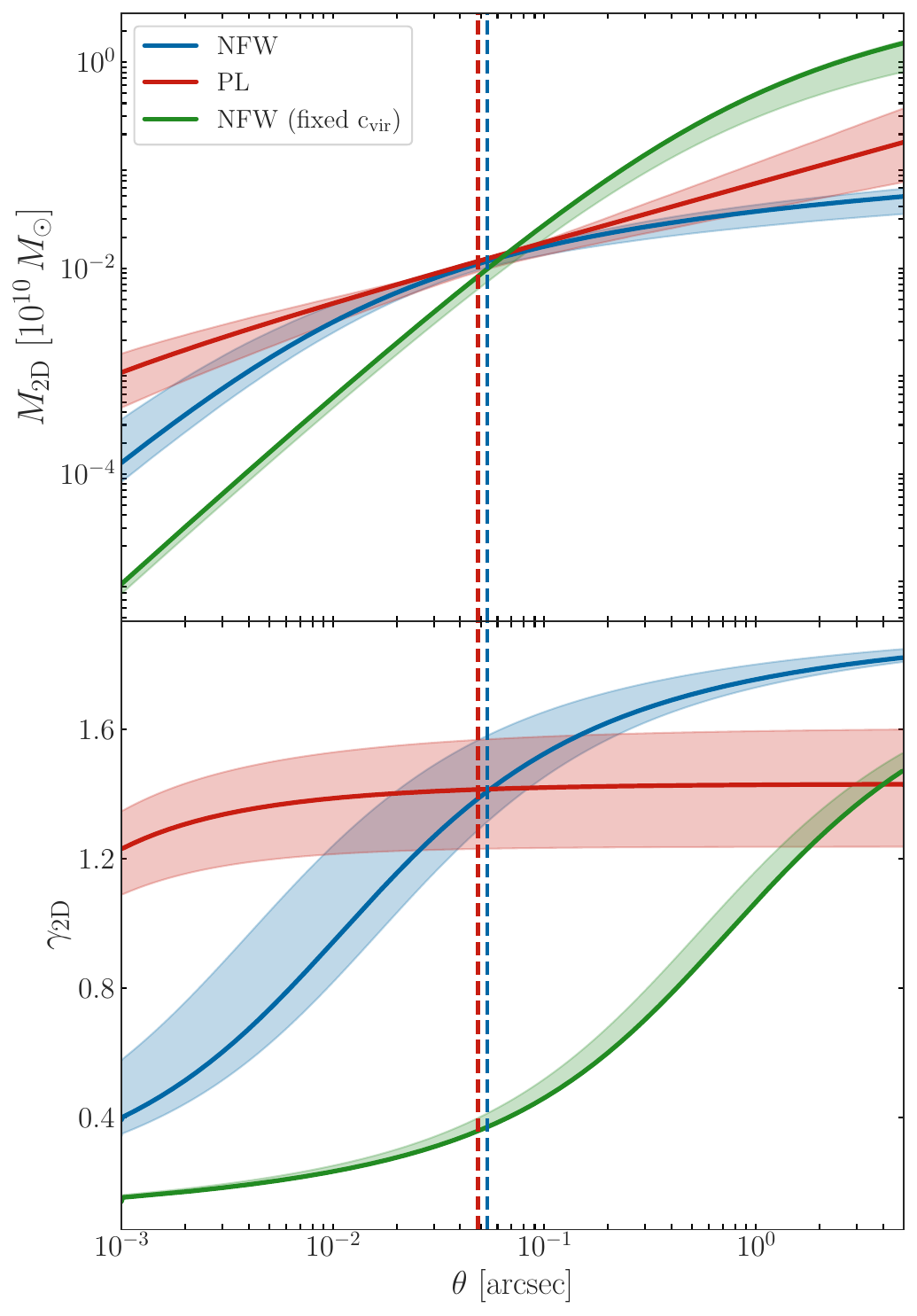}
\caption{Projected mass (top panel) and its logarithmic slope (bottom panel) profiles for the different models fitted to the perturber in the mock data. The dashed lines indicate the pivot radius for each model, in the corresponding colour.}
\label{fig:masses}
\end{figure}

Fig.~\ref{fig:masses} shows the inferred projected enclosed mass and logarithmic slope, $\gamma_{\rm 2D} \equiv 1 - c_1 = 2 - \tfrac{d \ln M_{2D}}{d \ln \theta}$, of the halo for all three models as a function of projected distance from its centre, with the pivot radii of the best-fitting models marked by vertical lines. The pivot radii of the preferred models are $\theta_0 = 0.0536$ arcsec for the free-concentration NFW and $\theta_0 = 0.0486$ arcsec for the PL model, separated by a scale much smaller than the PSF FWHM. Adopting the pivot radius of the best-fitting model as the reference, both models recover consistent projected enclosed masses and slopes at $\theta_0$, that agree with one another and with the true values to within $\sim 1\sigma$, $\log (M_{\rm 2D}^{\rm NFW}(\theta_0)/M_\odot)=8.07 \pm 0.05$, $\log (M_{\rm 2D}^{\rm PL}(\theta_0)/M_\odot)=8.09\pm 0.05$ and $\log (M_{\rm 2D}^{\rm truth}(\theta_0)/M_\odot)=8.13$; $\gamma_{\rm 2D}^{\rm NFW}=1.41 \pm 0.15$, $\gamma_{\rm 2D}^{\rm PL}=1.42 \pm 0.17$ and $\gamma_{\rm 2D}^{\rm truth}=1.31$. 
From Fig.~\ref{fig:masses}, it is clear that the inferred projected mass profiles of the two best-fitting models are consistent over a radial range around their pivot radii and diverge further from $\theta_0$. 
The disfavoured fixed-concentration NFW model, by contrast, agrees with neither of the two other models nor the truth in this radial range. The projected enclosed mass and slope at $\theta_0$ inferred from this model are $\log (M_{\rm 2D}^{\rm NFW (fix)}(\theta_0)/M_\odot)=7.99 \pm 0.07$ and $\gamma_{\rm 2D}^{\rm NFW (fix)}=0.37 \pm 0.03$, deviating from the truth at the $\sim 2 \sigma$ and $\sim 31 \sigma$ level, respectively. This follows directly from the reduced flexibility of this model compared with the other two. In this case, the model has a fixed radial shape with a single free parameter that can only rescale the amplitude of the deflection profile in one direction.

\section{Sensitivity to the perturber profile}
\label{sec:profile}

The formalism developed in Section \ref{sec:pivot_radius} allows us to estimate the errors on the local properties of the perturber profile within the assumed family of models. These precisions are, however, conditional on the fitted family being an adequate description of the true profile, which is not known a priori. 
Here, we ask how sensitive the strong lensing data are to local departures from the assumed two-parameter profile. This is essential to understand whether families of models with different local profiles can be compared in terms of their Bayesian evidence. Such differences, where detected, may then be indicative of features in the true underlying projected mass distribution of the perturber.

We start by considering a reference profile family described by two global parameters, $\boldsymbol{\eta}$. As in the previous section, $c_n(\bm{\eta})$ are the local expansion coefficients of the deflection profile around the pivot radius $\theta_0$. 
We then allow for an extended family, in which the base family is nested, with local perturbations $\delta c_n$ superimposed on the base coefficients, so that the general member of the family is described by $\{\bm{\eta}, \bm{\delta c}\}$. Setting $\boldsymbol{\delta c}= 0$ recovers the base family as a special case, while $\boldsymbol{\delta c} \neq 0$ spans the broader, augmented family that allows for local deviations from the base profile.
The logarithmic deflection angle profile about the pivot radius is then approximated as follows
\begin{equation}
  \ln\alpha_p(\theta;\, \boldsymbol{\eta},\boldsymbol{\delta c})
  = \sum_{n=0}^{N-1}\frac{c_n (\boldsymbol{\eta}) + \delta c_n}{n!}\,\ell^{\,n} \,,
  \label{eq:alpha_perturbed}
\end{equation}
where $N-1$ is the truncation order of the local expansion. 

We now want to determine, at each order $m$, how well the data can resolve a perturbation $\delta c_m$ to the perturber profile at the pivot radius. To this end, we evaluate the $(N+2)\times(N+2)$ Fisher matrix at the fiducial state, i.e., at $\boldsymbol{\delta c} = \boldsymbol{0}$, which in block form reads as follows
\begin{equation}
  \mathcal{I} = 
  \begin{pmatrix} 
    \mathcal{I}_{\boldsymbol{\eta}\boldsymbol{\eta}} & \mathcal{I}_{\boldsymbol{\eta}\boldsymbol{\delta c}} \\[0ex] 
    \mathcal{I}_{\boldsymbol{\delta c}\boldsymbol{\eta}} &  
    \mathcal{I}_{\boldsymbol{\delta c}\boldsymbol{\delta c}}
  \end{pmatrix}\,.
\end{equation}
The marginal Fisher information for the parameters of interest $\boldsymbol{\delta c}$ can then be calculated by marginalising out the global parameters $\boldsymbol{\eta}$, which is equivalent to taking the Schur complement of the $\mathcal{I}_{\boldsymbol{\eta}\boldsymbol{\eta}}$ block,
\begin{equation}
    \mathcal{I}^M_{\boldsymbol{\delta c}\boldsymbol{\delta c}} = \mathcal{I}_{\boldsymbol{\delta c}\boldsymbol{\delta c}} - \mathcal{I}_{\boldsymbol{\delta c}\boldsymbol{\eta}}\mathcal{I}_{\boldsymbol{\eta}\boldsymbol{\eta}}^{-1}\mathcal{I}_{\boldsymbol{\eta}\boldsymbol{\delta c}}\,.
\end{equation}
The diagonal elements, at the pivot radius, reduce to
\begin{equation}
\label{eq:i_deltac}
    [\mathcal{I}^M_{{\boldsymbol{\delta c}\boldsymbol{\delta c}}}]_{mm} = \frac{\mu_{2m}}{(m!)^2}-\frac{[\mathcal{I}]_{A\,\delta c_m}^{\,2}}{[\mathcal{I}]_{AA}}-\frac{[\mathcal{I}]_{s\,\delta c_m}^{\,2}}{[\mathcal{I}]_{ss}}\,.
\end{equation}
$\mathcal{I}^M_{\boldsymbol{\delta c}\boldsymbol{\delta c}}$ includes all the degeneracies between perturbation coefficients $\boldsymbol{\delta c}$ and the global parameters $\boldsymbol{\eta}$. Note that, in particular, $[\mathcal{I}^M_{{\boldsymbol{\delta c}\boldsymbol{\delta c}}}]_{00}$ vanishes identically for all parametrisations considered here, in which a zeroth-order perturbation is by construction equivalent to a rescaling of the amplitude parameter.

The first term in Eq.~\eqref{eq:i_deltac} is independent of how the family is parametrised; the information on a localised $m$-th order perturbation is encoded in the even moment $\mu_{2m}$ and falls rapidly with increasing order. This upper bound on the constraining power of the data is set by the arc geometry and its radial extent about the perturber, and constraining each successive order requires the pixels with the highest weights $w_k$ to span a progressively wider range of log-radii about the pivot radius.  The remaining two terms measure how much of a perturbation $\delta c_m$ can be reabsorbed by changing the global parameters $A$ and $s$, respectively. 
While the second term is common to all two-parameter models parametrised as in Eq.~\eqref{eq:alpha_two_reparam}, for which 
${[\mathcal{I}]_{A\,\delta c_m}^{\,2}}/{[\mathcal{I}]_{AA}} = \mu_m^2/\bigl((m!)^2\mu_0\bigr)$, the last term depends on the specific fitted model, through the parameter $s$. 

We now assume that the true perturber profile belongs to the augmented family, with parameters $\{\bm{\eta}, \bm{\delta c} \neq 0\}$, and investigate whether the data can statistically distinguish a base-family model from an augmented-family one on the basis of their Bayesian evidence.
Under the Laplace approximation, i.e., Taylor expanding the log-likelihood around its peak to second order, with the Fisher matrix as the (negative) Hessian,
the difference between the expected maximum log-likelihood of the base model ($\boldsymbol{\delta c}=0$), $L^\prime_0$, and that of the augmented model, $L_0$, is given by
\begin{equation}
\begin{split}
    -2\Delta\ln L ={}&
    (\bm{\eta} - \hat{\bm{\eta}})^T \mathcal{I}_{\bm{\eta}\bm{\eta}}
    (\bm{\eta} - \hat{\bm{\eta}})
    - (\bm{\eta} - \hat{\bm{\eta}}^\prime)^T \mathcal{I}_{\bm{\eta}\bm{\eta}}
    (\bm{\eta} - \hat{\bm{\eta}}^\prime) \\
    &- 2 (\bm{\eta} - \hat{\bm{\eta}})^T \mathcal{I}_{\bm{\eta}\bm{\delta c}}\, \hat{\bm{\delta c}}
    + \hat{\bm{\delta c}}^T \mathcal{I}_{\bm{\delta c}\bm{\delta c}}\, \hat{\bm{\delta c}}\,,
\end{split}
\label{eq:deltaloglike_nested}
\end{equation}
where $\Delta\ln L = \ln (L^\prime_0 / L_0)$, and $\hat{\bm{\eta}}^\prime$ and $(\hat{\bm{\eta}}, \hat{\bm{\delta c}})$ are the
maximum likelihood estimators (MLEs) for the parameters of the base and the augmented model, respectively.
Note that each model has its own maximum likelihood parameters.
Since the true profile belongs to the augmented family, the expected likelihood of the augmented model is maximised at the true parameter values.
Differentiating Eq.~\eqref{eq:deltaloglike_nested} with respect to ${\bm{\eta}}$ yields the difference between the MLE for the global parameters of each model,
\begin{equation}
    \Delta\bm{\eta} =
    \hat{\bm{\eta}}^\prime - \hat{\bm{\eta}} =
\mathcal{I}^{-1}_{\boldsymbol{\eta}\boldsymbol{\eta}}\,\mathcal{I}_{{\boldsymbol{\eta}\boldsymbol{\delta c}}}\,\hat{\bm{\delta c}}\,.
\label{eq:bias}
\end{equation}
Substituting Eq.~\eqref{eq:bias} in Eq.~\eqref{eq:deltaloglike_nested} then yields
\begin{equation}
    -2\Delta\ln L =
    \hat{\bm{\delta c}}^T\,\mathcal{I}^M_{{\boldsymbol{\delta c}\boldsymbol{\delta c}}}\,\hat{\bm{\delta c}}\,.
\end{equation}
For a given distortion $\hat{\delta c}_m$, with all other perturbation coefficients held at zero, this reduces to
\begin{equation}
    \big|2\Delta\ln L\big| =
    {\left(\frac{\hat{\delta c}_m}{\sigma_{\delta c_m}}\right)}^2\,,
    \label{eq:pert_snr}
\end{equation}
where $\sigma_{\delta c_m}=\sqrt{1/[\mathcal{I}^M_{{\boldsymbol{\delta c}\boldsymbol{\delta c}}}]_{mm}}$ is the forecasted error on $\delta c_m$, conditioned on the vanishing of all other perturbation coefficients. $\Delta\ln L$ is hence related to the signal-to-noise ratio of $\delta c_m$ in each dataset. 

The data therefore have, in principle, the statistical power to discriminate between the base and the augmented models through the Bayes factor, which trades off the gain in the maximum likelihood against the added model complexity according to Occam's razor \citep[see, e.g.,][for how the Bayesian evidence can be calculated directly from the Fisher matrix]{Heavens_2007}.
Distinguishing between the two models, however, requires the signal-to-noise ratio of $\delta c_m$ in the data to be large enough for the Bayes factor to decisively favour one model over the other.
Strong lensing data are thus not only sensitive to the projected enclosed mass of the low-mass perturbers within the robust radius, but in principle also to the higher-order local properties of their profiles around the same radius.

The resulting bias in the inferred projected enclosed mass at the pivot radius due to model misspecification can then be obtained from Eq.~\eqref{eq:bias} as
\begin{equation}
    \Delta \ln M_{\rm 2D}(\theta_0) =
    \frac{1}{\mu_0}\sum_{n=1}^{N-1}\frac{\mu_n}{n!}\,\hat{\delta c}_n\,.
    \label{eq:mass_bias}
\end{equation}
In particular, for a power-law model, $\mu_1=0$ at the pivot radius, so the leading term in the projected enclosed mass bias derives from a curvature mismatch between the base model and the true profile,
\begin{equation}
    \Delta \ln M_{\rm 2D, PL}(\theta_0) \approx
    \frac{\mu_2}{2\mu_0}\,\hat{\delta c}_2\,.
\end{equation}

Increasing the signal-to-noise ratio of the data through a longer exposure time, $t_{\rm exp}$, rescales all pixel weights by $t_{\rm exp}$, which cancels out in every ratio $\mu_n/\mu_0$ appearing in the sum of Eq.~\eqref{eq:mass_bias}. The systematic mass bias is therefore left unchanged, although the sensitivity to the local properties of the perturber profile is enhanced through the increase in $\mu_{2m}$, which encodes the information about a given $\delta c_m$. The angular resolution of the data, on the other hand, changes the distribution of weights in log-radius, and hence affects not only the sensitivity to the finer structures in the perturber profile, but also the accuracy of the mass measurement.

\section{Summary}
\label{sec:discussion}

We have introduced a data-driven robust scale for presenting marginalised constraints on the projected mass profile of low-mass perturbers in strong lensing analyses, which hold for both substructures in the main lens galaxy and field haloes along the line of sight.

The point of departure of our analysis is the Fisher information matrix, from which we have identified a pivot radius where the local amplitude and logarithmic slope of the profile decorrelate. This is, equivalently, the radius at which the projected enclosed mass of the profile is measured most precisely, and the definition applies to any sufficiently smooth deflection angle profile that can be expanded locally as a Taylor series. Although the pivot radius is model-dependent, models that are flexible enough to provide an adequate fit to the data recover nearly the same enclosed mass, within their statistical uncertainties, over a radial range around their very similar pivot radii. As a result, the pivot radius becomes a model-independent robust radius for mass measurements.

We note that the robust scale for mass estimates defined in this work differs structurally from the perturbation radius introduced by \citet{Minor_2017}. The latter is defined as the distance from the inferred subhalo position to the point on the critical curve that is perturbed the most, and was proposed as the radius within which the projected mass of the perturber can be robustly inferred. 
The perturbation radius is, therefore, determined purely from the geometry of the perturbed critical curve of the best-fit model and, unlike the definition proposed in this paper, does not account for the parameter uncertainties, and hence for the observational noise and data angular resolution, except through their effect on the best-fit values.

We have also shown that resolved strong gravitationally lensed emission has the statistical power to constrain the local properties of the projected mass density distribution of low-mass perturbers beyond their enclosed mass, and therefore to distinguish between different mass profiles. Constraining higher-order local features of the profile requires progressively more radial leverage around the perturber, as encoded in the higher moments, with a precision that depends on the SNR and angular resolution of the data. We conclude, therefore, that preferred models that provide a good fit to the data recover a consistent enclosed mass at the robust radius; the Bayesian evidence then discriminates between them through their ability to reproduce the local slope and, if the data provide sufficient radial leverage, the curvature and possibly higher-order features of the true profile.

Characterising the mass density profiles of the dark low-mass perturbers detected through strong gravitational lensing is essential to advancing our understanding of the nature of dark matter, as it enables direct comparison with the theoretical predictions of different dark matter models. The framework presented in this work identifies the scale over which such comparisons can be made in a model-independent way.

\section*{Acknowledgements}
The authors are grateful to Simon White for helpful discussions and feedback. This research was carried out on the High-Performance Computing resources of the Freya cluster at the Max Planck Institute for Astrophysics in Garching, operated by the Max Planck Computing and Data Facility. S.~V. thanks the Max Planck Society for support through a Max Planck Lise Meitner Research Group.

\section*{Data Availability}

The data underlying this paper will be shared on reasonable request to the corresponding author.




\bibliographystyle{mnras}
\bibliography{ms}



\appendix

\bsp	
\label{lastpage}
\end{document}